\documentclass[a4paper,12pt]{article}

\usepackage{amsmath}
\usepackage[final]{graphicx}
\usepackage{bm}
\usepackage{amssymb}

\usepackage{cancel}

\usepackage{amssymb}
\newcounter{comment}

{\refstepcounter{comment}%
\begin{quote}
\ttfamily\small$\blacksquare$ \textbf{\underline{Comment} $\sharp$\thecomment:}}%
{\end{quote}}

\begin{document}
\hfill

\begin{center}
\baselineskip=2\baselineskip
\textbf{\LARGE{Critique of Fermionic R$\nu$MDM and its
Scalar Variants}}\\[6ex]
\baselineskip=0.5\baselineskip

{\large Kre\v{s}imir~Kumeri\v{c}ki,
Ivica~Picek
and
Branimir~Radov\v{c}i\'c
}\\[4ex]
\begin{flushleft}
\it
Department of Physics, Faculty of Science, University of Zagreb,
 P.O.B. 331, HR-10002 Zagreb, Croatia\\[3ex]
\end{flushleft}
\today \\[5ex]
\end{center}

\begin{abstract}

We examine the stability of minimal dark matter (MDM) particle-candidates
in the setup in which they participate in radiative neutrino (R$\nu$) masses.
We first point out the existence of an additional renormalizable term in recently proposed R$\nu$MDM Lagrangian, which violates the claimed accidental $Z_2$ symmetry and spoils the stability of the fermionic MDM quintuplet component.
We then explore the viability of R$\nu$MDM variants based on scalar MDM multiplets. There are ubiquitous super-renormalizable terms in the scalar potential which make these scalar multiplets unstable.

\end{abstract}

\vspace*{2 ex}

\begin{flushleft}
\end{flushleft}

\clearpage

\section{Introduction}

There are two experimental pieces of evidence \cite{Nakamura:2010zzi} pressing us to extend the standard model (SM): the evidence for neutrino masses and for the dark matter (DM) content of the Universe. The ongoing experiments at the Large Hadron Collider (LHC) may provide us with new insights on new heavy particles enabling Weinberg's dimension-five operator $LLHH$ \cite{Weinberg:1979sa} as an explanation of the smallness of masses of observed neutrinos.

The known tree level realizations of  Weinberg's operator proceed by adding to the particle content of the SM a single, GUT-scale multiplet: a heavy fermion singlet for the type-I \cite{Minkowski:1977sc-etc}, a scalar triplet for the type-II \cite{KoK-etc} and a fermion triplet for the type-III \cite{Foot:1988aq} seesaw mechanism.
On the other hand additional discrete symmetries are instrumental \cite{Ma:2000cc,Grimus:2009mm,Xing:2009hx} in bringing the seesaw mechanism to the TeV scale accessible at the LHC.

Alternative lowering of the seesaw scale, without discrete symmetries, can be achieved  either
by going beyond dimension-five operator or by generating the neutrino masses at the loop level. The approach used by two recent models is to employ non-zero hypercharge seesaw mediators:
the dimension-seven mechanism with fermion triplets  in \cite{Babu:2009aq} and the dimension-nine mechanism with fermion quintuplets in \cite{Picek:2009is,Kumericki:2011hf}. Both of these mechanisms offer the spectacular triply-charged states verifiable at the LHC.

The approach in \cite{Ma:2006km}, which we follow here, is to link a  DM candidate to the scale of radiatively generated neutrino (R$\nu$) masses. The authors of \cite{Cai:2011qr} exemplify it by employing   a 10 TeV  scale
fermionic quintuplet of zero hypercharge. Let us stress that an isolated $\Sigma \sim (5,0)$ fermion field or an isolated septuplet $\Phi \sim (7,0)$ scalar field have been singeld out as viable DM particle candidates within the so-called minimal dark matter (MDM) model \cite{Cirelli:2005uq}. The dubbed  R$\nu$MDM model \cite{Cai:2011qr} claims to achieve  the stability of the neutral MDM quintuplet component without introducing a symmetry beyond the SM gauge symmetries, on account of an accidental $Z_2$ symmetry arising from the choice of the field content. It has been adopted in more
recent paper \cite{Chen:2011bc} aimed at successful leptogenesis.

From the point of generating the neutrino masses, the fermion quintuplet has to be accompanied by an appropriate scalar multiplet. High enough scalar field multiplet \cite{Cai:2011qr} avoids the terms in the scalar potential which may lead to and induced vacuum expectation value (vev) of the new scalar field, and there are only radiative neutrino masses. This R$\nu$MDM is realized by accompanying the zero hypercharge quintuplets
$\Sigma \sim (5,0)$ by a hypercharge one sextuplets,  $\Phi \sim (6,1)$, proposing the neutral component $\Sigma^0$ as
fermionic DM candidate.

In Section 2 we give a brief description of this attempt and explicate
a previously omitted term in the scalar potential,
which spoils the stability of $\Sigma^0$ particle.
In Section 3 we analyze  two  new seesaw models, the first in which a tentative DM particle appears as a neutral component of the scalar quintuplet $\Phi \sim (5,0)$, accompanied by a quadruplet seesaw mediator $\Sigma \sim (4,1)$. In the second model we show that originally viable MDM scalar septuplet $\Phi \sim (7,0)$ candidate also becomes unstable in the
context of generating radiative neutrino masses.  We summarize our results in the concluding Section.

\section{Instability of fermionic quintuplet R$\nu$MDM}
\label{fermipetplet}

\begin{figure}[h]
\centerline{\includegraphics[scale=1.0]{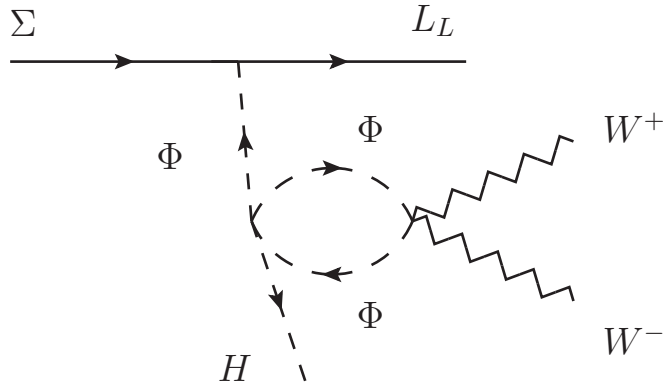}}
\caption{\small An example of decays of the heavy lepton $\Sigma$ at the loop level.}
\label{decay}
\end{figure}

To allow a Yukawa coupling of the isospin two fermion $\Sigma \sim (5,0)$ to the lepton doublet $L_L \sim (2,-1) $, necessary for a mass generation mechanism,
\begin{equation}\label{lagrangian}
   \mathcal{L}_Y = \overline{L_L} Y \Phi  \Sigma  + \mathrm{H.c.} \ ,
\end{equation}
one has to introduce a scalar field $\Phi$ of a half-integer (3/2 or 5/2) isospin in addition to the SM higgs $H \sim (2,1)$. In order to avoid a vev of $\Phi$ which might come from quartic $H^3 \Phi$ term, the fermionic quintuplet $\Sigma \sim (5,0)$ has to be accompanied by a sextuplet scalar $\Phi \sim (6,-1)$ as in \cite{Cai:2011qr}. These authors have argued that the neutral component of the fermionic quintuplet, $\Sigma^0$, can be a DM candidate on account of an accidental $Z_2$ symmetry of the Lagrangian. We show that it is a viable DM candidate only if one forbids or finetunes a renormalizable $H \Phi^3$ term with the quartic coupling $\lambda$, omitted in \cite{Cai:2011qr}. In their tensor notation, the omitted term reads
\begin{equation}\label{dim4}
\lambda \Phi^* \Phi^* \Phi H^* + \mathrm{H.c.} \ \ \ , \ \ \ \Phi^* \Phi^* \Phi H^* =
\Phi^{*iabcd} \Phi^{*pqrst} \Phi_{abpqr} H^{*n} \epsilon_{in} \epsilon_{cs} \epsilon_{dt} \ ,
\end{equation}
and it can induce a rapid decay of a DM particle at the loop level. For example, the loop diagram on Fig. \ref{decay} generates an effective dimension-six operator
\begin{equation}
    \mathcal{O}_6 = \overline{L_L}^i \Sigma_{jklm} W^l_a W^m_b H^{*n} \epsilon^{ja} \epsilon^{kb} \epsilon_{in}\ ,
\end{equation}
which leads to decays of the DM candidate $\Sigma^0$. In particular, we estimate the decay amplitude for  $\Sigma^0 \rightarrow \nu W^+ W^- H$ arising from it,
\begin{equation}
    A \sim \frac{g^2}{16\pi^2} Y \lambda \frac{m_\Sigma}{m_\Phi^2} \frac{m^2_\Sigma}{m^2_W}   \ .
\end{equation}
It produces the decay width into four final-state particles,
\begin{equation}
    \Gamma \sim \frac{1}{2 m_\Sigma} \ |A|^2 \  d\mathrm{LIPS_4}  \, ,
\end{equation}
which is given by
\begin{equation}
    \Gamma \sim \frac{g^4}{192 \pi} \Big(\frac{1}{16\pi^2}\Big)^4 Y^2 \lambda^2 \frac{m_\Sigma^9}{m_\Phi^4 m^4_W } \, .
\end{equation}
The inverse of it, a lifetime of a DM candidate, should be longer then the age of the Universe given by the Hubble time $H^{-1} \approx 10^{17} {\rm s}$. For $\Sigma^0$ to have the lifetime $\tau_{\rm DM} > 10^{17} {\rm s}$, the quartic coupling $\lambda$ has to be tiny. By taking $g=0.65$, $m_\Sigma=m_\Phi=10 \ \rm{TeV}$ and $Y=10^{-1}$, the values adopted from \cite{Cai:2011qr}, we obtain
\begin{equation}
    \lambda < 10^{-20}
\end{equation}
as an upper limit. In the context of the decaying DM \cite{Valle} the bound on the $\tau_{DM}$ can be nine orders of magnitude larger ($\tau_{\rm DM} > 10^{26} {\rm s}$), implaying even stronger limit, $\lambda < 10^{-24}$.
This means that $\Sigma^0$ is a viable DM candidate only if the coupling in Eq.~(\ref{dim4}) has an extremely small value. This is in contradiction with the claim in \cite{Cai:2011qr}, and in order to avoid the fine-tuning one has to impose the $Z_2$ symmetry forbidding  the renormalizable term in Eq.~(\ref{dim4}), by hand.

Without such discrete symmetry a model more minimal than in \cite{Cai:2011qr} is possible with scalar quadruplet $\Phi \sim (4,-1)$ replacing previous sextuplet field. Like in case of previously studied non-zero hypercharge fermion \cite{Picek:2009is,Kumericki:2011hf}, this field $\Phi$ develops an induced vev from the quartic $\Phi H^3$ term. The Yukawa terms result in a tree-level dimension-nine seesaw operator. These and additional contributions to the neutrino masses from radiative loop-suppressed diagrams are presented in a separate paper \cite{KPR12_Ferm5-pletY0}.

\newpage

\section{R$\nu$MDM Variants with Scalar MDM}
\label{scalarpetplet}

\begin{figure}[h]
\centerline{\includegraphics[scale=1.3]{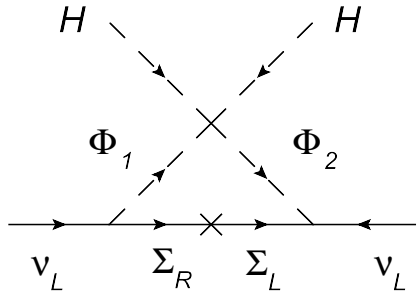}}
\caption{\small Diagram for radiatively generated neutrino masses where one of the scalar fields contains the MDM component.}
\label{mnuloop}
\end{figure}

In the light of the described blow to fermionic MDM candidate in the radiative neutrino masses setup, let us now reconsider the scalar MDM multiplets \cite{Cirelli:2005uq,Hambye:2009pw,Hally:2012pu}. The models involving hypercharge-zero scalar field in radiatively generating neutrino masses through diagrams displayed on Fig.~\ref{mnuloop} have two essential features: (i) the scalar field is accompanied with appropriate fermion multiplet; (ii) there are two scalar fields of different hypercharge in play, the one with hypercharge zero being the DM candidate.

\subsection{Model with scalar quintuplet DM}

A minimal possibility to realize the diagram on Fig.~\ref{mnuloop} through the scalar quintuplet $\Phi_1 \sim (5,0)$ as MDM candidate is to accompany it with another scalar quintuplet $\Phi_2 \sim (5,2)$ in conjunction with non-zero hypercharge fermionic quadruplet $\Sigma \sim (4,1)$ as the seesaw mediator.

\begin{figure}[h]
\centerline{\includegraphics[scale=1.0]{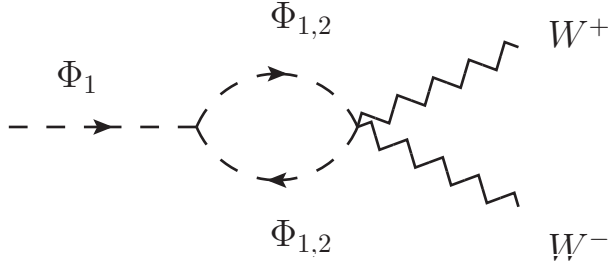}}
\caption{\small An example of the loop diagram for quintuplet $\Phi_1$ decays.}
\label{decay2}
\end{figure}

With two scalars $\Phi_1 \sim (5,0)$ and $\Phi_2 \sim (5,2)$, there are, up to our knowledge previously unconsidered, dimension-three $Z_2$ noninvariant operators in the Lagrangian
\begin{equation}\label{scalar5}
    \mathcal{L}_S = \mu_1 \Phi_1 \Phi_1 \Phi_1 + \mu_2 \Phi_1 \Phi_2 \Phi^*_2 \ ,
\end{equation}
which in tensor notation read
\begin{eqnarray}\label{dim31}
\nonumber
  \Phi_1 \Phi_1 \Phi_1 &=& \Phi_{1ijkl} \Phi_{1mnpq} \Phi_{1rstu} \epsilon^{im} \epsilon^{jn} \epsilon^{kr} \epsilon^{ls} \epsilon^{pt} \epsilon^{qu} \ ,\\
  \Phi_1 \Phi_2 \Phi^*_2 &=& \Phi_{1ijkl} \Phi_{2mnpq} \Phi_2^{*klpq} \epsilon^{im} \epsilon^{jn}\ .
\end{eqnarray}
These terms make the DM candidate $\Phi_1^0$ unstable through loop diagrams like the one on Fig.~\ref{decay2}. We estimate the amplitude for the decay $\Phi_1^0 \to W^+ W^-$
\begin{equation}
    A \sim \frac{g^2}{16\pi^2} (\mu_1 + \mu_2) \frac{m^2_{\Phi_1}}{m^2_W}
\end{equation}
and the decay width
\begin{equation}
    \Gamma \sim \frac{g^4}{16 \pi} \Big(\frac{1}{16\pi^2}\Big)^2 \ (\mu_1 + \mu_2)^2 \ \frac{ m^3_{\Phi_1}}{ m^4_W} \ .
\end{equation}
To have the lifetime of a DM candidate $\tau_{\rm DM} > 10^{17} {\rm s}$, the parameters $\mu_{1,2}$ in Eq.~(\ref{scalar5}) have to have very small values, $\mu_{1,2} < 10^{-10} {\rm eV}$.

\subsection{Model with scalar septuplet DM}

The second possibility to realize the diagram on Fig.~\ref{mnuloop} is through the scalar septuplet $\Phi_1 \sim (7,0)$ as MDM, together with another scalar septuplet $\Phi_2 \sim (7,2)$, now in conjunction with fermionic sextuplet $\Sigma \sim (6,1)$.

For a single scalar septuplet $\Phi_1$, the term $\Phi_1^3$ is forbidden by Bose statistics \cite{Hambye:2009pw}, but in presence of another septuplet $\Phi_2$ there is the dimension-three $Z_2$ noninvariant operator
\begin{eqnarray}\label{dim3}
\nonumber
  \mathcal{L}_S &=& \mu \Phi_1 \Phi_2 \Phi^*_2 \ ,\\
  \Phi_1 \Phi_2 \Phi^*_2 &=& \Phi_{1ijklmn} \Phi_{2pqrstu} \Phi_2^{*lmnstu} \epsilon^{ip} \epsilon^{jq} \epsilon^{kr} \ ,
\end{eqnarray}
making the DM candidate $\Phi_1^0$ unstable.
\begin{figure}[h]
\centerline{\includegraphics[scale=1.0]{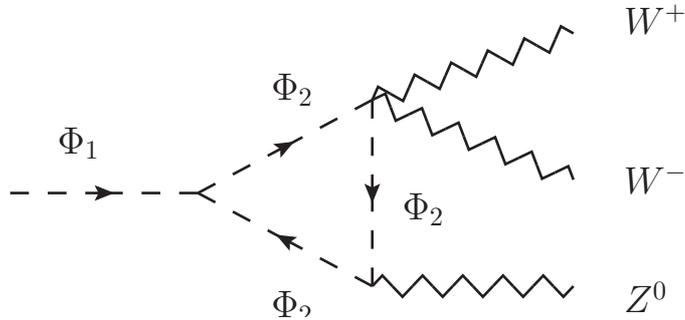}}
\caption{\small An example of the loop diagram for septuplet $\Phi_1$ decays.}
\label{decay3}
\end{figure}
These terms make the DM candidate $\Phi_1^0$ unstable through loop diagrams, like the one on Fig.~\ref{decay3}. We estimate the amplitude for the decay $\Phi_1^0 \to W^+ W^- Z^0$
\begin{equation}
    A \sim \frac{g^3}{16\pi^2} {\mu \over m_{\Phi_1}} \frac{m^3_{\Phi_1}}{m^3_W} \ ,
\end{equation}
and the decay width
\begin{equation}
    \Gamma \sim \frac{g^6}{32 \pi} \Big(\frac{1}{16\pi^2}\Big)^3 \ \mu^2 \ \frac{ m^5_{\Phi_1}}{ m^6_W} \ .
\end{equation}
To have the lifetime of a DM candidate $\tau_{\rm DM} > 10^{17} {\rm s}$, the parameter $\mu$ in Eq.~(\ref{dim3}) is restricted to very small value, $\mu < 10^{-11} {\rm eV}$.

To conclude, a major blow to scalar R$\nu$MDM variants comes from super-renormalizable dimension-three $\Phi^3$ operators. Besides these renormalizable $Z_2$ violating operators, there are possible effects from the Planck scale considered in \cite{Cirelli:2005uq,Valle} and references therein.

\section{Conclusions}

We address in a common framework the open questions of the neutrino masses and the dark matter content of the Universe. In view of the immense DM possibilities, a minimality of the DM model appears as a criterion essential for its predictivity and testability. The minimality is satisfied by adding only one extra $SU(2)_L$ multiplet in the MDM setup. However, by imposing additional seesaw mission to a given MDM multiplet, it has to be accompanied by additional scalar field introducing new obstacles to the DM stability. It would be appealing if the DM candidate, a neutral component of a certain higher weak-isospin multiplet, could be stabilized without introducing a symmetry beyond the SM gauge symmetries, as claimed in \cite{Cai:2011qr}.
We have shown that this is not possible for already selected MDM particles, if the neutrino masses are generated through simple one-loop diagrams on  Fig.~\ref{mnuloop}. The obstacle for the fermionic quintuplet MDM candidate in \cite{Cai:2011qr} comes from renormalizable quartic term omitted there. Concerning the scalar R$\nu$MDM variants, we have shown that the stability of the scalar quintuplet and septuplet MDM candidate is violated by  ubiquitous super-renormalizable terms that can be tamed only by imposing extra discrete symmetry.

\end{document}